# Two-Dimensional Water Diffusion at a Graphene-Silica Interface

DaeEung Lee, Gwanghyun Ahn and Sunmin Ryu*

Department of Applied Chemistry, Kyung Hee University, Yongin, Gyeonggi 446-701, Korea

ABSTRACT: Because of the dominant role of the surface of molecules and their individuality, molecules behave distinctively in a confined space, which has far-reaching implications in many physical, chemical and biological systems. Here, we demonstrate that graphene forms a unique atom-thick interstitial space that enables the study of molecular diffusion in 2-dimensions with underlying silica substrates. Raman spectroscopy visualized intercalation of water from the edge to the center underneath graphene in real time, which was dictated by the hydrophilicity of the substrates. In addition, graphene undergoes reversible deformation to conform to intercalating water clusters or islands. Atomic force microscopy confirmed that the interfacial water layer is only *ca*. 3.5 angstroms thick, corresponding to one bilayer unit of normal ice. This study also demonstrates that oxygen species responsible for the ubiquitous hole doping are located below graphene. In addition to serving as a transparent confining wall, graphene and possibly other 2-dimensional materials can be used as an optical indicator sensitive to interfacial mass transport and charge transfer.

## Introduction

Capillary action governing everyday phenomena, such as absorption of water-soluble nutrients by plants, the flow of tears through tear ducts in eyes and the wetting of soil by rain drops, occurs along the interfaces of continuous bulk water and surfaces of different solid materials. The unique behavior of water in such a constrained space, originating from the rising importance of molecular interactions at a smaller length scale, plays an important role in diverse fields, such as biology, geology, meteorology and nanotechnology.[1-3] Even more intriguing properties have been revealed for water molecules confined to nanoscopic dimensions, such as nanopores and nanotubes. Aquaporin, a membrane pore proteins, achieve selectivity for water molecules over protons via an ingenious molecular gate, permitting passage only to those that satisfy the preset hydrogen bonding pattern.[4] Despite the apparent enthalpic penalty due to the loss of hydrogen bonds, water spontaneously fills carbon nanotubes (CNTs)[5] as a result of entropic gains,[6] and its flow rate is a few orders of magnitude higher than that expected in the continuum hydrodynamic limit.[7] Because of the subtle balance between water-water and water-wall interactions, the ice-liquid transition of water entrapped in CNTs is dictated by their diameters.[6,8]

Intercalation in the interplanar 2-dimensional space of graphite has a longer history of research than the 1-dimensional channel in the nanotubes regarding charge-transfer binding and superconductivity.[9] Additionally, graphite intercalated with alkali metals is an indispensable electrode material for rechargeable batteries and reducing agents.[9] The free energy stabilization required for the molecular transport into a nanoscopic space can be achieved either through decreasing enthalpy or increasing entropy. The contributions of the two driving factors depend significantly on the geometric and chemical structures of the confining walls as shown in the simulated counter-intuitive water filling in CNTs.[5,6] Scanning probe microscopy revealed the importance of the attractive interaction between the intercalants and the walls in water diffusion under graphene along hydrophilic mica[10] and Ru(0001)[11] that forms stable water bilayers.[12] Particularly, extremely hydrophilic mica was found to maintain ordered water layers in ambient conditions when covered with graphene.[13-15] Despite the apparent hydrophobicity[16] of graphene and reduced hydrogen bonding, theory also predicted that the weak van der Waals (vdW) interaction between water and C atoms stabilizes a monolayer of water confined between a pair of graphene sheets.[17,18] In this regard, the Janus-like[19] interface consisting of hydrophobic graphene and hydrophilic silica can serve as an interesting model system for water, an intercalant bearing the significant implications.[1-3] The hydrophilicity of the silica surface can be systematically tuned by modifying surface functional groups[20] or charge balance.[21]

In studying mass transport in a two-dimensional confined space, graphene supported on a solid substrate can provide a unique and unprecedented opportunity due to its unique geometry and various properties. Ultrathin,[22] flexible[23] and yet mechanically robust[24] graphene membrane can serve as not only a confining wall but also a window that is transparent for electronic[25,26] and optical[27,28] probes in wide energy and wavelength ranges, as recently reported by several research groups: bulk-like staging behavior in the intercalation of halogen molecules[29] and alkali metals[30] through few-layer graphene, dissociative adsorption-driven intercalation of oxygen through graphene-Ru(0001),[25,31] water-intercalation-induced splitting of graphene on Ru(0001)[11] and electron microscopy of CO intercalation through a graphene-Pt(111) interface.[32] The complete[33] or partial[34] transparency of graphene towards "wetting" may allow an additional control on the behavior of the intercalants by adsorbing a third molecular film on top of the graphene. In addition, the intercalant-induced bending and deformation,[35] which may activate the Raman D band,[36] can be exploited for molecule-level understanding of the intercalation process.

Despite the high optical transparency[27,28] and Raman spectroscopic sensitivity[36] of graphene, however, optical microscopy of mass transport behavior under graphene has not been reported. Optimal design of graphene electronic devices also demands understanding possible interfacial diffusion of water and its effects on the device performance, since typical graphene electronic devices are supported on hydrophilic oxide dielectrics and exposed to ambient humidity. In addition, better understanding of the interface may enable us to fully explain the widely studied spontaneous charge doping occurring at the graphene-silica systems.[37-43] While the surface[37,38] or interfacial[39,42,43] oxygen species originating from the ambient air have been proposed as the universal hole dopants, the mechanistic details have yet to be revealed. In particular, the role of interfacial water[44] in the charge transfer doping is far from being understood because of the chemical complexity of the silica surface.

In this study, we report an unprecedented real-time Raman microscopic observation of water intercalation through a Janus-faced 2-dimensional space confined within a sub-nm by hydrophobic graphene and hydrophilic silica walls. When annealed graphene in direct contact with silica was submerged, intercalation of water occurred gradually from the graphene edges at a rate of <3 µm/hour and led to a *ca.* 3.5-angstrom-thick interfacial water layer, as confirmed by atomic force microscopy (AFM). Intercalation-induced reversible deformation of the lattice was confirmed by modulation in the D band intensity. The intercalation could be accelerated by the introduction of nanopores serving as an additional entrance on graphene basal planes. The optical contrast generated by the changes in the Raman G and 2D frequencies of graphene was modulated by hole-doping oxygen species, which were removed by water intercalants. This result also illuminates the mechanism of oxygen-induced charge transfer doping of graphene.

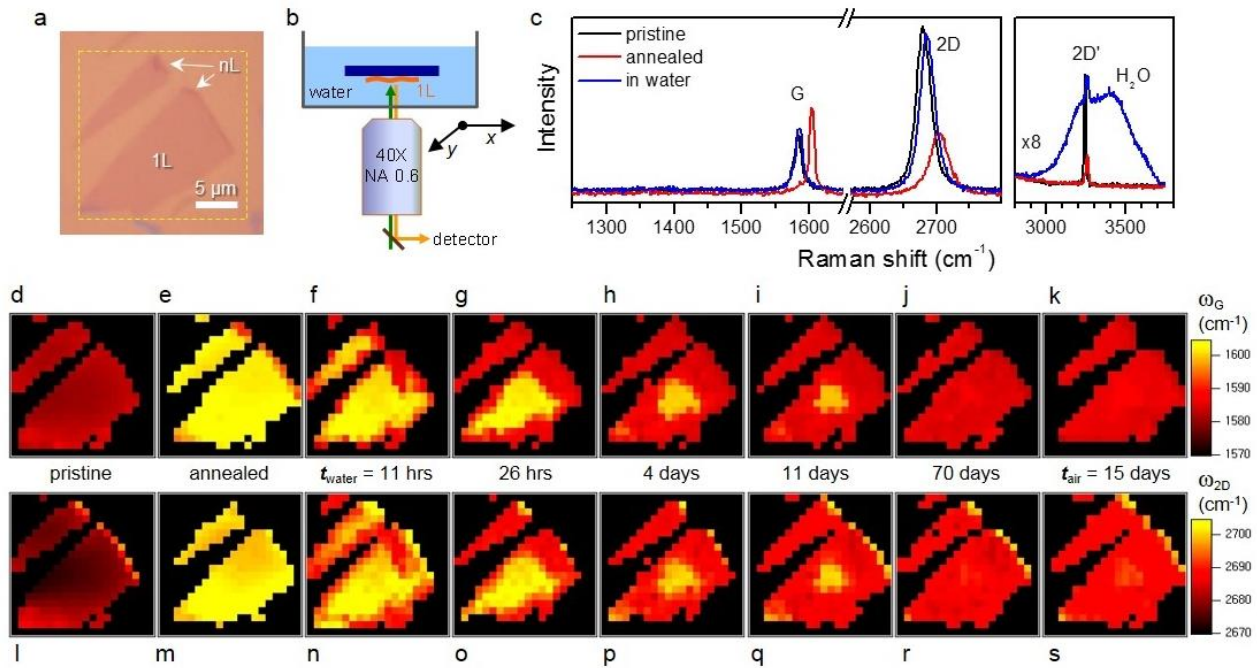

Figure 1. Raman microscopy of the edge-to-center spectral recovery caused by water diffusion. (a) Optical micrograph of the 1L sample D1. Small multilayer (nL) regions are found at the edges. The dashed square represents an area (15 x 15 µm²) raster-scanned during Raman mapping. (b) A simplified scheme for in-situ Raman measurements. See Fig. S1 for detailed geometry of the employed water cell. (c) Representative Raman spectra sets obtained for pristine, annealed and submerged graphene samples: (left) sample D1; (right) sample D2. Both were annealed at 400 °C for 2 hrs in a vacuum. The immersion time ($t_{water}$) was 70 days and 12 hours for D1 and D2, respectively. (d–j) $\omega_G$-Raman maps obtained for D1 in its pristine, annealed and submerged state for varying immersion time. Each pixel represents an area of 1 x 1 µm². (k) $\omega_G$-Raman map of D1 re-exposed to air for 15 days. (l–r) $\omega_{2D}$-Raman maps obtained for D1 in its pristine, annealed and submerged state. (s) $\omega_G$-Raman map of D1 re-exposed to air for 15 days.

Results

The single layer (1L) graphene sample (D1) in Fig. 1a was mechanically exfoliated from graphite onto $SiO_2$/Si substrates and annealed at 400 °C to induce oxygen-mediated hole doping[37,39], which provides an optical contrast as explained below (see Supporting Information Section A for detailed methods). The Raman spectra in Fig. 1c reveal that the annealing treatment up-shifts the G and 2D Raman peak frequencies ($\omega_G$ and $\omega_{2D}$, respectively) and attenuates the intensities of the 2D and 2D' peaks ($I_{2D}$ and $I_{2D'}$), primarily because of the charge transfer doping and lattice contraction as explained below.[45] However, when immersed in water (Fig. 1b) at room temperature for a prolonged period (note the water peak denoted as $H_2O$ obtained from D2 in Fig. 1c), the Raman spectrum of graphene was nearly completely recovered to that of pristine graphene. To obtain spatial information on the water-induced change, Raman mapping was performed in situ as a function of the immersion time ($t_{water}$). Figure 1d reveals that $\omega_G$ is relatively constant and close to its intrinsic value of 1581 cm$^{-1}$ throughout the entire pristine 1L area, except for the multilayer (nL) area near the edges (Fig. 1a). The $\omega_G$-map in Fig. 1e confirms that the annealing-induced upshift of $\Delta\omega_G$ ~ 20 cm$^{-1}$ shown in Fig. 1c is also spatially uniform, except for the multilayer (nL) area near the upper edges (Fig. 1a). When submerged, however, the recovery of $\omega_G$ observed in Fig. 1c occurred from the graphene edges (Fig. 1f for $t_{water}$ = 11 hours) and proceeded gradually to the central area of D1 (Fig. 1h–1j), with completion occurring at $t_{water}$ = 70 days. When D1 was removed from the water cell right after obtaining Fig. 1j and exposed to the ambient air during the subsequent 15 days, no back-upshift in $\omega_G$ was observed as shown in Fig. 1k. The $\omega_{2D}$-maps in Fig. 1l–1s also show an identical recovery pattern from the edges. Whereas the rate of change and time for completion varied significantly from sample to sample, an identical trend was confirmed for other samples (Supporting Information Section B). Notably, graphene decoupled from the substrates with the water layers remained intact without scrolling or detachment during the prolonged observation periods.

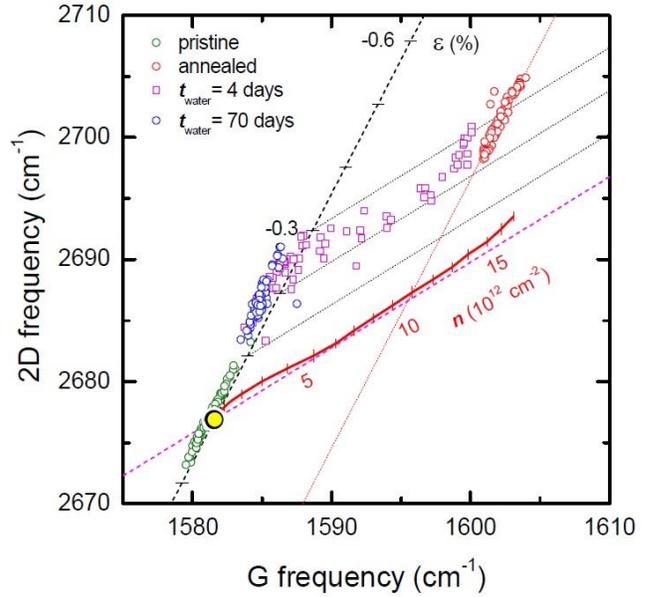

Figure 2. Decomposition of the effects of charge density ($n$) and mechanical strain ($\varepsilon$). A portion of the Raman map data in Fig. 1 are presented in the $\omega_G$-$\omega_{2D}$ space. The yellow circle represents the intrinsic Raman frequencies of graphene, O (1581, 2677) cm$^{-1}$.[45] Hole doping and lattice contraction ($\varepsilon$ < 0) will shift the point along the $n$-axis (magenta dashed line) and $\varepsilon$-axis (black dashed line), respectively. The red solid line along the $n$-axis represents the experimental data of Das et al.[45,46]

To quantify the changes in the charge density ($n$) and the mechanical strain ($\varepsilon$) in D1, the vector analysis proposed by Lee et al.[45] was employed for the Raman map data shown in Fig. 1. In the $\omega_G$-$\omega_{2D}$ correlation graph (Fig. 2), the point moves from the origin O ($\omega_G°$, $\omega_{2D}°$), representing charge-neutral and unstrained graphene, along the $n$- or $\varepsilon$-axis, in response to hole doping or uniaxial strain, respectively. Because the effects of the two are independent of each other, any given ($\omega_G$, $\omega_{2D}$) can be vector-decomposed to give $n$ and $\varepsilon$ values.[45] This result reveals that the spectral variation for the pristine D1 (green circles) is due to mechanical strain (-0.1% < $\varepsilon$ < 0.1%), and its native charge density is very small ($n$ < 10$^{12}$ cm$^{-2}$). It can also be observed that the annealing-induced upshifts in $\omega_G$ and $\omega_{2D}$ (red circles) are the combined effects of hole doping and lattice contraction, as previously reported.[45] Despite the overall in-plane compression (-0.35% < $\varepsilon$ < -0.2%), the spread in strain is largely maintained ($\Delta\varepsilon$ ~ 0.15%), and the degree of charge doping is quite homogeneous, $n$ = (1.0 ± 0.05) x 10$^{13}$ cm$^{-2}$. The data points for D1 submerged for 4 days, however, are scattered in parallel with the $n$-axis, which indicates that the charge density varies widely (0 < $n$ < 9 x 10$^{12}$ cm$^{-2}$) with the $\varepsilon$ maintaining a rather constant value (-0.3% to -0.2%). The value of $n$ is low for the edge regions and high for the center (Fig. 1h & 1p). For the case of $t_{water}$ = 70 days (blue circles), all the data points are aligned on the $\varepsilon$-axis, which reveals that the water-induced recovery in $\omega_G$ and $\omega_{2D}$ is essentially due to the undoping that was initiated from the edges. Note that this conclusion is also supported by other spectral features, such as the linewidths of G ($\Gamma_G$) and 2D ($\Gamma_{2D}$) and the $I_D/I_G$ ratio (see Supporting Information Section C).

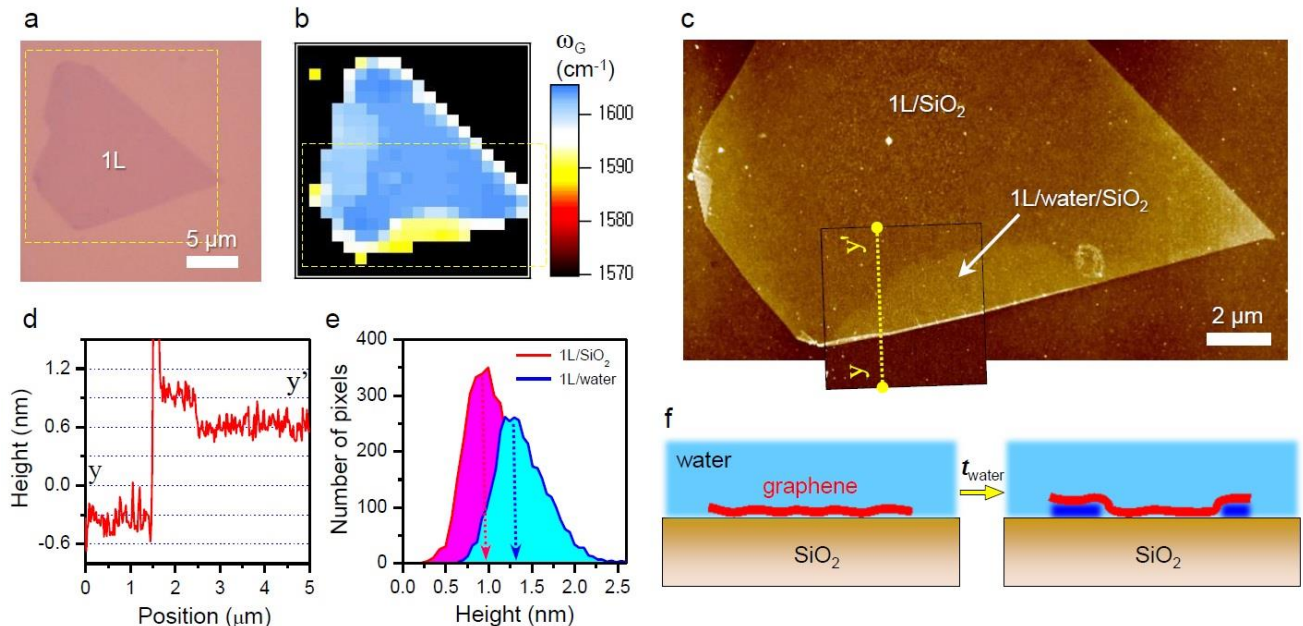

Figure 3. Topographic confirmation of interfacial water layers. (a) Optical micrograph of the 1L sample D3. The dashed square indicates the region in which the Raman map (b) was obtained. (b) The $\omega_G$-Raman map of D3 obtained in the air after 14 hour-submersion following thermal annealing at 400 °C for 8 hours. (c) Non-contact AFM height images obtained from the region in the dashed rectangle of B after the Raman mapping. The plateau in the lower portion of D3 is due to an interfacial water layer. Three images (two that are 10x10 µm²; one that is 5x5 µm²) were stitched for better visibility. (d) The height profile along the y-y' line (c), which shows a 0.35-nm-high step between 1L/SiO₂ and 1L/water/SiO₂. (e) Height histograms for the 1L/SiO₂ and 1L/water/SiO₂ regions obtained from the 5 x 5 µm² image. The difference in the average heights (vertical dotted arrows) agrees with the height of the step of d. (f) Schematic representation of water intercalation from the edge to the center through the graphene-silica interface.

We found that the spectral recovery is induced by intercalation of water through the graphene-silica interface. To uncover morphological changes of graphene in water using AFM, another 1L sample (D3 in Fig. 3a) was annealed at 400 °C and submerged for 14 hours. The $\omega_G$-Raman map in Fig. 3b reveals that the water-induced spectral recovery occurred near the edges, particularly in the bottom edge, with most of the central region unchanged. In the AFM height image (Fig. 3c) obtained from the lower part of D3, there is a clearly flat plateau (~8 x 2 µm²) located along the bottom edge. Remarkably, the plateau matches well with the yellow-colored area in the Raman map in position, shape and size. The height profile across the plateau (Fig. 3d) reveals that there is a step of *ca.* 0.35 nm in height. A statistical analysis using height histograms in Fig. 3e is also consistent with the presence of a molecular layer. By conducting two cycles of submersion and AFM measurement for another sample, we confirmed that water-induced plateaus expand in area as increasing $t_{water}$ (see Supporting Information Section D). Based on the spatial match between the spectroscopic and topographic maps, we conclude that the plateau is formed by intercalation of water through the graphene-silica interface, as depicted by the schematic diagram in Fig. 3f. Since the SiO₂ substrates are amorphous and have protrusions on the atomic scale,[47] the graphene-silica interface is unlikely to induce atomically flat water films that were observed on the crystalline substrates.[14,15,48] However, its average height comparable to that of "puckered bilayers" in normal hexagonal ice ($I_h$)[13] suggests that the trapped water is essentially one-molecule-thick, possibly one bilayer, which will be further discussed below. The edge-to-center recovery further unveils that oxygen molecules, responsible for the observed hole doping, are not located on but underneath graphene[39] and are to be replaced by water intercalants during the recovery process, as will be discussed below.

By generating a dense array of nanopores in the basal plane of graphene, completion of intercalation could be accelerated significantly. To form nanopores, sample D4 in Fig. 4a was partially oxidized at 550 °C following the work of Liu et al.[37] The AFM image in Fig. 4b shows a random array of nanopores with a diameter of 50 ± 10 nm and a density of 11 ± 3 µm⁻². As Liu et al. showed,[37] oxidation induced upshifts in $\omega_G$ and $\omega_{2D}$ (Fig. 4c & 4d), which were attributed to the same hole doping as observed in graphene annealed in a vacuum. Figure 4g confirms that the post-oxidation spectral changes are due to hole doping and compression, identical to the case of D1 annealed in a vacuum (Fig. 2). With the nanopores, however, the spectral recovery of D4 in water was significantly faster. As shown in Fig.4g, $n$ decreased by ~80% within the first 2 hours and became nearly charge-neutral in 9 hours, which contrasts with the long period (> 11 days) required for D1 without nanopores (Fig. 1 & 2). In addition, undoping by water occurred rather uniformly across the entire area of D4 instead of selectively from the outer edges, as observed for D1. These results substantiate that water diffuses through the nanopores, which serve as additional intercalation channels.

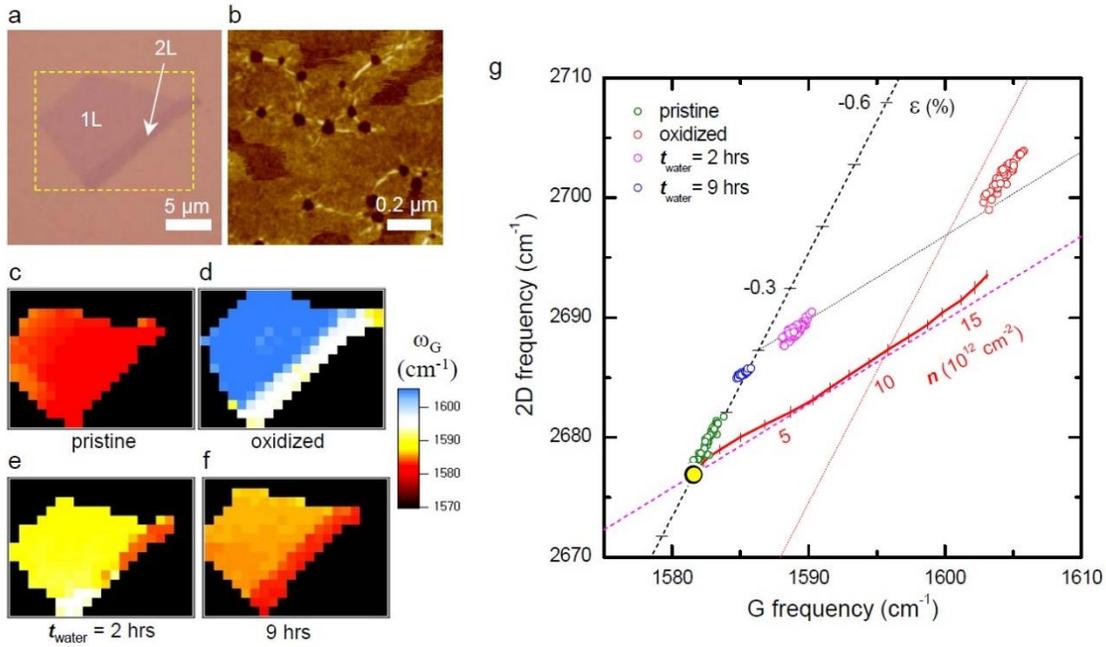

Figure 4. Enhanced mass flow through the nanopores. (a) Optical micrograph of the 1L sample D4. The dashed rectangle indicates the region in which the Raman maps (c ~ f) were obtained. (b) AFM image of the 1L area of D4, which was oxidized at 550 °C for 30 min to generate nanopores with a diameter of ~50 nm. (c–f) The $\omega_G$-Raman maps obtained for D4 in its pristine, oxidized and submerged states, respectively. (g) The Raman map data presented in ($\omega_G$, $\omega_{2D}$) space.

We further revealed that the water intercalation leads to reversible deformation of the graphene lattice. Figure 5a presents a series of Raman spectra obtained from D5 as a function of $t_{water}$. To accelerate water intercalation, nanopores were generated in D5 through oxidation, which also induced hole doping. As water intercalated, the G peak exhibited clear asymmetry at $t_{water}$ = 15–17.5 min and finally downshifted to ~1585 cm$^{-1}$ at $t_{water}$ = 20 min, confirming undoping (see also Fig. 5b for the $I_{2D}/I_G$ ratio). More remarkably, the intercalation also modulated the D peak intensity. With increasing $t_{water}$, the $I_D/I_G$ ratio in Fig. 5b reached a maximum of 0.23 simultaneously with complete undoping ($t_{water}$ = 20 min) and then decreased to the original value (~0.05), which was caused by the edges of the nanopores ($t_{water}$ = 40 min).[37] The D peak is generated through a double resonance process only when electrons scattered by a D phonon with a large wave vector are scattered back to the original position in the Brillouine zone by structural defects.[49] Carbon atoms at the nanopore edges serve as these scattering defects because they lack translational symmetry of the lattice. The increase in the $I_D/I_G$ ratio followed by the complete recovery indicates that the intercalation of water induces reversible disorders, not permanent bond breaking. We conclude that the graphene steps at the water fronts, as visualized in the AFM image (Fig. 3c), are serving as efficient scattering defects. Because nanometer scale ripples in graphene and highly curved graphene in CNTs with ~1 nm diameter exhibit negligible $I_D$ values ($I_D/I_G$ ~0.01),[37,50], the water-induced steps are likely to have a higher curvature, possibly resulting in a deformation on the atomic length scale. In D5, there could be a significant number of such water fronts that are propagating in a radial direction from each nanopore. Non-uniform intercalation should impose tensile stress on the graphene membrane. Strain-charge density analysis showed that the membrane underwent tensile deformation ($\Delta\varepsilon$ ~0.1% at $t_{water}$ < 15 min), which was relieved simultaneously with the recovery of the $I_D/I_G$ ratio (Supporting Information Section E). The time delay between the complete recovery of charges ($t_{water}$ ~ 20 min) and disorders ($t_{water}$ ~ 40 min) suggests that the initial growth of the interfacial layers may be more complex than a simple layer-by-layer growth pattern. We attribute the reversible activation of the D peak to its high conformability due to elastic out-of-plane deformation and substantial adhesion to substrates.[47] In this regard, Dimiev et al. recently reported a similar observation for sulfuric acid intercalating through bulk graphite.[35]

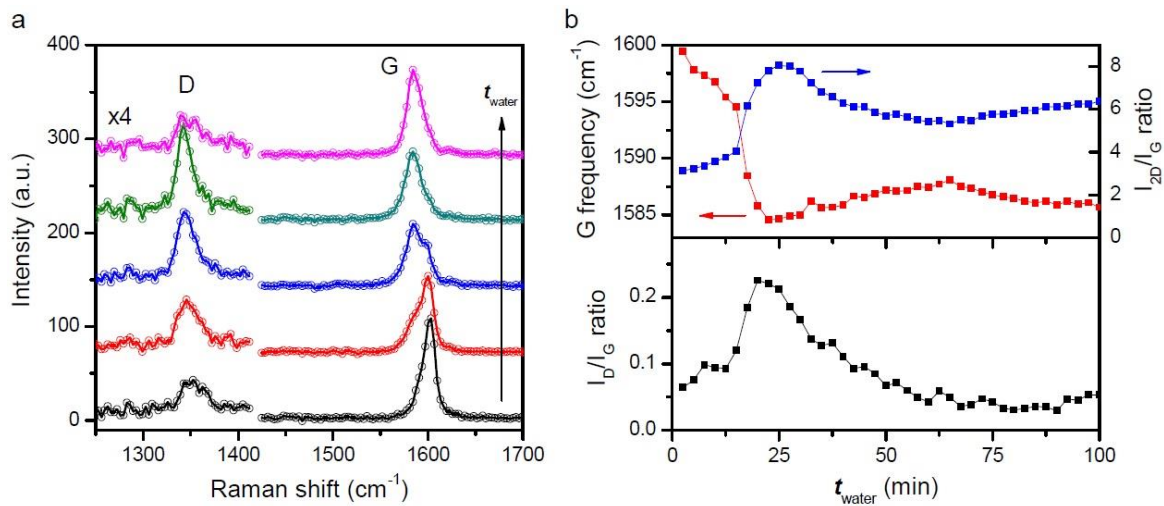

Figure 5. Reversible lattice deformation during water intercalation. (a) Raman spectra obtained from the submerged 1L graphene (D5): $t_{water}$ = 2.5, 15, 17.5, 20 and 45 min (from bottom to top). To expedite water intercalation, nanopores were generated in the basal plane through oxidation at 550 °C for 30 min as in D4 of Fig. 4. (b) (Top) The G frequency (red) and the $I_{2D}/I_G$ ratio (blue) as a function of $t_{water}$. Because of the asymmetry of the G peak, the first moment of the G peak was used as a representative value for $\omega_G$. (Bottom) The $I_D/I_G$ ratio as a function of $t_{water}$.

We also found that the interfacial water behaves very differently for highly hydrophilic substrates. To modify hydrophilicity represented by the water contact angle, bare substrates were treated with piranha or $O_2$ plasma. In Fig. 6a and 6b, the contact angle decreased from 45° for the as-received substrates to 23° and 3° for piranha- and plasma-cleaned ones, respectively. The high hydrophilicity of plasma-treated silica is due to surface charges.[21] Because adsorption of ambient gaseous hydrocarbon contaminants gradually increases the contact angle [16] (Fig. S6), all the measurements were completed within 10 min after each cleaning. Figure 6c shows another 1L sample (D6), which was prepared on a plasma-treated substrate. When submerged as prepared for ~1 s, the entire graphene sheet of D6 turned into a scroll (Fig. 6d), suggesting that graphene near the edges was detached from the substrates and rolled into the scroll, minimizing the free energy of the system[51] (see Fig. 6g). As observed in the AFM image in Fig. 6e, the diameter of the scroll, assumed to have a circular cross section, was 80–100 nm. A simple arithmetic scroll model (see Supporting Information Section G) predicts that the interlayer distance of the scroll is 0.5–1.1 nm, which suggests that the scroll is tightly wound and may contain, if any, one or two monolayers of water in the interlayer space at most. Scrolling or ensuing complete detachment of 1L graphene from the plasma-treated substrates was so highly efficient that only one sample out of 58 survived the submersion (see Fig. 6f). However, as the thickness increased, the survival probability increased rapidly to ~15% for 3L and ~85% for thicker flakes (>15L) as a result of increasing elastic bending energy required for scrolling or folding.[52] For pristine samples prepared on as-received or piranha-cleaned substrates, however, no change was observed during the prolonged one-month submersion, revealing the dominating role of hydrophilicity in the interfacial diffusion of water.

Discussion

Interfacial diffusion and effects of interfacial hydrophilicity

Our study visualized water fronts propagating underneath graphene through direct topographic imaging and Raman microscopy. The initial diffusion rate of water varied from 0.1 to 3 μm/hour among samples, indicating high spatial inhomogeneity of the factors that govern the spontaneous interfacial mass flow. Nanoscopic mass transport is significantly affected by the dimensions and chemical nature of the confining walls.[5,8,17] Filling of water through CNTs, for example, is driven by either enthalpic or entropic gain, depending on their diameter, which dictates the local hydrogen bonding structure of water molecules.[6]

As shown in Fig. 6, we note that among the key variables is hydrophilicity of the silica surface, which is determined primarily by the surface charge density[21] or the relative population of silanols and siloxanes.[20] Since we do not currently have the capability to determine the local chemical nature of the silica surface, the water contact angle instead served as a macroscopic indicator for hydrophilic moieties. When the substrate surface was made strongly hydrophilic by O$_2$ plasma, the pristine graphene was either scrolled or detached within seconds of submersion. In contrast, the graphene on less hydrophilic substrates (as-received or piranha-cleaned) exhibited no noticeable morphology changes upon the prolonged submersion. This observation underlines the dominant role of hydrophilicity of substrates in the interfacial diffusion of water.

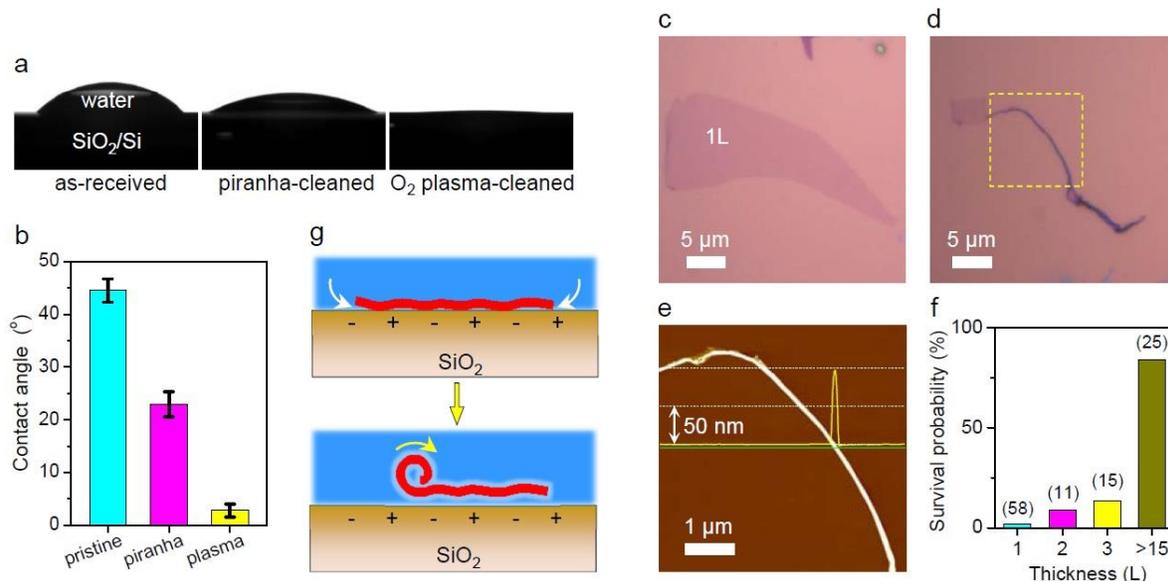

Figure 6. Interfacial water-induced scrolling of graphene on hydrophilic substrates. (a) Photographs of water drops on SiO$_2$/Si substrates with different prior treatments for contact angle measurements: (left) as-received, (middle) cleaned with a piranha solution, (right) cleaned with O$_2$ plasma. (b) Water contact angles on the three substrates in a. (c) Optical micrograph of the 1L sample D6 prepared on plasma-cleaned substrates. (d) Optical micrograph of the scrolled D6 obtained after submersion for 1 s in water. (e) Height AFM image of D6 obtained from the dashed square in d. The height profile (yellow solid line) obtained along the green line reveals that the diameter of the scroll is ~100 nm. (f) Survival probability against scrolling or detachment. The duration of immersion was less than 3 seconds. The numbers of samples tested are noted in the parentheses in the bar graphs. All of the samples were deposited on plasma-cleaned substrates. (g) Schematic diagrams for water-induced scrolling: (top) intercalation of fluidic water layers between highly hydrophilic substrates and graphene; (bottom) thermally activated scrolling of submerged graphene, which is completely surrounded by water.

Hindered scrolling or detachment

Although graphene sheets were completely surrounded by water in the submerged graphene/water/substrate systems in Fig. 1 and Fig. S2, they remained intact in water up to several months, which contrasts with the rapid scrolling or detachment of graphene supported on highly hydrophilic substrates as shown in Fig. 6. The prolonged stability of D1 suggests that the interfacial water is not fluidic but forms a rigid layer that serves as a solid-like substrate to hold graphene in place through vdW interactions.[17] It is this graphene-water layer interaction that compensates for the slight loss of hydrogen bonding experienced by the water monolayer confined between two graphene sheets.[17] This explanation is also supported by the presence of ice-like water held by immobilized silanols on silica surfaces in ambient conditions, as confirmed by vibrational spectroscopy methods.[3,53] It was also predicted that a rigid monolayer water on crystalline silica is stable up to 300 K.

In contrast, the substrates treated with O$_2$ plasma led to immediate scrolling or detachment of graphene upon submersion, which suggests that water intercalation occurred extremely fast. The increase in the rate can be attributed to the enhanced hydrophilicity which should provide enthalpic gain for the intercalation. It is to be noted that ambient water vapor readily intercalates through the interface between graphene and extremely hydrophilic mica substrates.[10] Based on the facile delamination of graphene, we conjecture that the plasma-treated substrates induce liquid-like interfacial water layers which fail to maintain stable adhesion of graphene upon submersion. In this regard, we note that the high affinity towards water of the plasma-treated substrates may result in thicker water layers that were observed on the mica substrates.[14] Unlike the first rigid water layer that is bound to the polar mica, the X-ray reflectivity[54] and scanning tunneling microscopy[14] revealed that the second and above layers are less ordered or liquid-like and thus highly susceptible to the external perturbation.

Once graphene is freed from the substrate by complete surrounding by liquid water, scrolling is strongly favored to minimize the free energy of the system,[52] as seen in the aggregation of reduced graphene oxide in aqueous media.[55] Considering their high hydrophilicity and significant amount of ambient water vapor, however, we cannot exclude the possibility that liquid water layer may form on the plasma-treated substrates even before deposition of graphene.

Location of hole dopants

The current study provides a direct evidence that the hole dopants are mainly located between graphene and silica, revealing an important aspect of the ubiquitous but elusive chemical interaction between carbon materials and ambient gases, particularly oxygen and water molecules. Spontaneous hole doping of graphene, first reported by Novoselov and Geim,[55] was later found to be uncontrollable and attributed to ambient gases or charge traps on the substrates.[56] Despite recent findings, such as annealing-induced amplification of the hole doping,[37] confirmation of $O_2$ as a hole dopant[39] and several related studies,[40,41,43,57,58] its detailed mechanism has yet to be uncovered, with a similar controversy remaining for CNTs.[59,60] Despite their clear sensitivity toward $O_2$ with[61] and without[59] metal contacts, CNTs bind $O_2$ through physisorption with negligible charge transfer and low binding energy, which would allow only one $O_2$ per $10^5$ C atoms under ambient conditions.[60] Even full charge transfer would lead to $n = $ ~$4 \times 10^{10}$ cm$^{-2}$, which is far lower than that observed in this study. Although formation of endoperoxides on CNTs was proposed in low pH solutions,[62] this explanation was discarded for graphene based on the absence of the D peak.[37] Our study also disproves the two mechanisms because physisorption and endoperoxide formation would not prefer the bottom surface of graphene to the top surface.

It should be noted that our study does not necessarily disprove the physisorption of $O_2$ on the top surface of graphene. In particular, graphene with thermally induced structural deformation[39] may provide an enhanced binding for $O_2$ and other ambient molecules, as was suggested in other studies.[40,41,43,57,58] We also note that all the measurable Raman spectroscopic changes underwent the edge-to-center undoping dynamics upon submersion as shown if Fig. 1, which cannot be expected for dopants physisorbed on the top surface of graphene. Thus we conclude that the charge density solely due to physisorbed $O_2$ would be very small fraction of what was observed in the current study.

Charge transfer doping and its reversal by intercalation

Our results are consistent with the charge transfer doping model involving a redox couple of $O_2/H_2O$, which explained hole doping of hydrogenated diamond surfaces in the ambient air.[63] By borrowing 4 electrons from graphene, $O_2$ can be reduced to four OH$^-$ ions consuming two $H_2O$ molecules. Most significantly, this scenario satisfies the thermodynamic requirement that the effective electron affinity[63] of the redox couple is larger than the work function[64] of graphene, which is consistent with the observed spontaneous hole doping. The availability of the redox couple in the ambient conditions suggests that the charge transfer reaction may occur in many other systems that satisfy the thermodynamic criterion.

Our study also demonstrated the presence of the hole dopants underneath graphene, indicating non-negligible mass flow underneath graphene despite the apparently precise seal at the graphene-silica interface.[65] Whereas graphene is generally hydrophobic, the silanol groups of the silica surface can attract water molecules,[3] which then participate in the electrochemical reactions. The annealing-induced nano-ripples[39] locally suspended[66] off from the substrate may provide highly efficient transport channels.

Assuming that 4 electrons are consumed by each redox couple,[63] $n = (1.0 \pm 0.05) \times 10^{13}$ cm$^{-2}$ of annealed D1 in Fig. 2 suggests that at least $2.5 \times 10^{12}$ $O_2$ molecules existed per cm$^2$ under the annealed graphene. The proposed electron transfer from graphene to the couple will generate OH$^-$ ions underneath cationic graphene, thus forming a dipolar interface. The hydroxide anions with an areal density that equals the hole density of graphene may be further stabilized by water molecules[37,39,43,57] provided from the ambient air. Upon submersion, however, more water intercalates from graphene edges as confirmed by the AFM measurements. The Raman maps suggests that the rigid water layers drive out the dopants by filling the vdW gap, which reverses the charge transfer. Despite the clear correlation, however, the back charge transfer process requires further explanation on its driving force and a detailed molecular picture.

Open questions

Despite the above novel findings on water intercalation, the current study leaves a few questions open for further investigation. Firstly, it needs to be understood how exactly water molecules are transported from edges to center. It is unlikely that the water layer collectively moves towards the center. Instead, numerous tiny water channels may be formed in the radial direction, as seen in dewetting of graphene/mica system.[10] The lack of such a fractal structure as found on mica[10] in our AFM images may be due to high roughness of the silica substrates which limited the visibility of fine geometrical details. Secondly, the role of geometrical gap between graphene and substrates as an intercalation entrance also deserves further study. Despite the overall intimate vdW contact, there is evidence[66,67] that graphene is locally suspended across atomically rough silica substrates providing loose vdW gap that may allow significant room for molecular motion. The observed wide distribution in the intercalation rate may be not only due to inhomogeneous local hydrophilicity but also to spatial variation in size of the vdW gap. Thirdly, the electrochemical reactions occurring in the nanoscopic space are far from being understood. Because of their ubiquitous nature and high relevance to general low dimensional systems, the interfacial doping and undoping processes deserve further theoretical and experimental investigations for molecule-level understanding.

In conclusion, we report real-time optical imaging of water diffusion in an atom-thick 2-dimensional space defined by graphene and silica substrates. Optical contrast was made by the interfacial hole dopant, $O_2$, which was removed by water molecules that diffuse in gradually from the graphene edges. Nanopores serving as additional entrances in the basal plane accelerated completion of the diffusion. The rate of diffusion could be significantly increased by increasing the hydrophilicity of the substrates. AFM revealed that the height of the interfacial water films is ~0.35 nm, which would correspond to only one bilayer of water in hexagonal ice. The diffusion-induced increase in the D peak intensity followed by a decrease upon completion of diffusion revealed that the graphene membrane undergoes severe and yet reversible deformation due to water intercalants. This study demonstrates that graphene and possibly other 2-dimensional materials can serve as ideal model systems for a nanoscopic confined space, which will lead to more discovery.

## ASSOCIATED CONTENT

Supporting Information. Methods, sample-dependent kinetics of water intercalation, additional spectral features of D1: $\Gamma_G$, $\Gamma_{2D}$ & $I_{2D}/I_G$, AFM characterization of moving water fronts, reversible deformation caused by intercalation of water, effects of ambient contaminants on water contact angles of bare substrates, estimation of spacing in graphene nanoscrolls, and supporting references. This material is available free of charge via the Internet at http://pubs.acs.org.

## AUTHOR INFORMATION

Corresponding Author

*E-mail: sunryu@khu.ac.kr

## ACKNOWLEDGMENT

This work was supported by a grant from the Kyung Hee University in 2009 (KHU-20090749) and also by the National Research Foundation of Korea (NRF-2011-0031630, NRF-2012R1A1A2043136). The authors thank Prof. Min Hyung Lee for the access to the $O_2$ plasma cleaner.

TOC Graphic

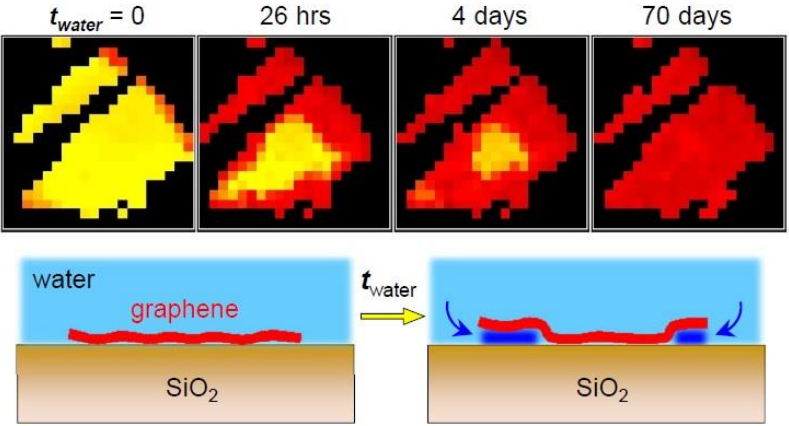